\documentstyle[eqsecnum,aps]{revtex}
\begin{document}

\thispagestyle{empty}
{\baselineskip-4pt
\font\yitp=cmmib10 scaled\magstep2
\font\elevenmib=cmmib10 scaled\magstep1  \skewchar\elevenmib='177
\leftline{\baselineskip20pt\vbox to0pt
   { {\yitp\hbox{Osaka \hspace{1.5mm} University} }
     {\large\sl\hbox{{Theoretical Astrophysics}} }\vss}}

\rightline{\large\baselineskip20pt\rm\vbox to20pt{
\baselineskip14pt
\hbox{OU-TAP-27} 
\vspace{2mm}
\hbox{February 19,1996}\vss}}%
\vskip3cm
\begin{center}{\large 
Gravitational waves from a spinning particle in circular orbits 
around a rotating black hole}
\end{center}
\vspace{0.5cm}
\begin{center}
 {Takahiro Tanaka, 
 Yasushi Mino\footnote{JSPS junior fellow}, Misao Sasaki and Masaru Shibata
 } \\
\vspace{0.5cm}
{\em Department of Earth and Space Science, \\
Osaka University,~Toyonaka 560,~Japan}\\
\vspace{0.4cm}
\end{center}

\begin{abstract}

Using the Teukolsky and Sasaki-Nakamura formalisms for 
the perterbations around a Kerr black hole, we calculate 
the energy flux of gravitational waves induced by 
a {\it spinning} particle of mass $\mu$ and spin $S$ 
moving in circular orbits near the equatorial plain of a 
rotating black hole of mass $M (\gg \mu)$ and spin $Ma$. 
The calculations are performed by using the recently developed 
post-Newtonian expansion technique of the Teukolsky equation. 
To evaluate the source terms of perturbations caused by a {\it spinning} 
particle, we used the equations of motion of a spinning particle 
derived by Papapetrou and the energy momentum tensor of a 
spinning particle derived by Dixon. 
We present the post-Newtonian formula of the gravitational wave luminosity 
up to the order $(v/c)^5$ beyond the quadrupole formula including 
the linear order of particle spin.  
The results obtained in this paper will be an important guideline to 
the post-Newtonian calculation of the inspiral of two spinning compact 
objects. 
\end{abstract}

\section{Introduction}

One of the most promising sources of gravitational waves for
kilo-meter size laser-interferometric detectors such as LIGO\cite{ligo}, 
VIRGO\cite{virgo} and future laser-interferometric detectors in space 
such as LISA\cite{lisa} is the coalescing compact binary of 
neutron stars and/or black holes.  
Since it is a highly general relativistic event, 
detection of gravitational waves from those binaries will bring us very 
fruitful information about relativistic astrophysical objects 
if we know physics of the final phase of the coalescence. 
Thus, there have been continuous efforts made by many authors to understand 
this phase\cite{will,kip}. 

Recently, it has been recognized that detection of the 
signal from a binary in the inspiraling phase 
is particularly very important because it can tell us a variety of 
parameters of the binary, i.e., mass, spin, 
etc.\cite{Cutler,cutl}. Furthermore it may provide some knowledge 
about the cosmological parameters\cite{Cutler,mar}.
In order to extract these important parameters from the data,
we need an accurate theoretical template of the wave form. 
Especially, the accumulated phase of the emitted gravitational 
waves is very sensitive to the binary parameters. 
Thus, the rate of change in the frequency of the orbital rotation 
due to the radiation reaction of gravitational waves 
must be evaluated accurately. 

The post-Newtonian expansion is the standard method to 
calculate the wave form of gravitational waves. 
Recently, the energy loss rate to 2PN order beyond 
the quadrupole formula has been derived by Blanchet et al.\cite{blan}
and to 2.5PN order by Blanchet\cite{blan2}
for the binary composed of non-spinning compact bodies.
The leading order effect of spin, which appears at 1.5PN and 2PN orders,
have been evaluated by
Kidder, Will and Wiseman\cite{kw2} and Kidder\cite{kidd}.

However, dificulties and complications increase exponentially as one
goes to higher orders with the standard post-Newtonian calculation technique.
Hence it will be very useful if we have different approaches to the 
higher order approximation and if we are able to provide some guiding
principle prior to the standard post-Newtonian calculations. 
The black hole perturbation approach is the 
only one known method that is independent of the standard
post-Newtonian approach and that can handle (not all but some important
portion of) higher order post-Newtonian effects in a relatively 
straightforward way.
In this approach, we consider a particle orbiting a black hole and
assume the mass of the black hole $M$ is much larger than that of the
particle $\mu$.

The black hole perturbation approach is based on the perturbation
equation derived by Teukolsky\cite{Teu}, which applies to a general
rotating (Kerr) black hole. One of great advantages of this approach is that
it takes full account of relativistic effects by construction and
numerical methods can be easily implimented to treat very general orbits.
Moreover, it has been shown that one can formulate an analytical
post-Newtonian expansion scheme in this approach as well. 
Poisson first noticed this fact and calculated the energy flux in the
case of a particle in circular orbits around a non-rotating black hole
to 1.5PN order\cite{pois}. A technically important point was to deal with
the Regge-Wheeler equation, which is equivalent to the Teukolsky
equation but has a much nicer property than it.
Along this line, Sasaki developed a systematic method to proceed to the 
higher orders in the case of a non-rotating black hole\cite{Sas}, 
and Tagoshi and Sasaki\cite{TagSas} 
gave the analytical expression for the energy flux up to 4PN order.
Meanwhile Poisson calculated the case of a rotating black hole
to 1.5PN order\cite{pois2}, but directly dealing with the Teukolsky
equation. Hence it seemed formidable to go beyond this order.
Then extending the method of ref.\cite{Sas},
a better method was developed by Shibata, Sasaki, Tagoshi and 
Tanaka\cite{SSTT} to treat the case of a rotating black hole
and the energy flux up to 2.5PN order was calculated.
This was made possible by using the Sasaki-Nakamura
equation\cite{SasNak}, which is
a generalization of the Regge-Wheeler equation for a non-rotating black
hole. Recently the calculation has been extended to 4PN order 
by Tagoshi et al.\cite{SSTT2}. 

In all of these previous papers, the small mass particle 
was assumed to be spinless. However, apart from the interest in its own
light, for the purpose of providing a better theoretical template
or at least a better guideline for higher
order post-Newtonian calculations, it is desirable to take 
into account the spin of the particle.
To incorporate this effect, we must know 
the energy momentum tensor of a spinning particle as well as
the equations of motion.
Fortunately, we can find them in the literature. The equations of motion
of a spinning particle were first derived by Papapetrou\cite{papa}, 
and they were put into more elaborate form by Dixon\cite{dixon} 
and Wald\cite{wald}. In particular, Wald\cite{wald} clarified all the conserved
quantities along the general particle trajectory.
On the other hand, Dixon\cite{dixon} succeeded to give the general form of
the energy momentum tensor of a particle with multipole moments,
which of course includes that of a spinning particle as a limit.
Hence, by using them, we can calculate the wave form and the 
energy flux of gravitational waves by a spinning particle 
orbiting a rotating black hole.

Here, a word of caution is appropriate. Usually, we regard the small
mass particle to be a model of a neutron star or a black hole.
However, if we regard the spinning small mass particle as a Kerr
black hole with mass $\mu$ and spin angular momentum $\mu S$,
it should have the definite multipole moments ($\propto \mu S^{\ell}$).
Since we neglect the contribution of these higher multipole moments 
in this paper, the particle in our treatment is not an adequate model 
for a (rapidly rotating) Kerr black hole. 
To incorporate the contribution of all higher multipole moments 
to represent a Kerr black hole is a future issue.
Here, we concentrate on the leading order effect due to the spin of 
the small mass particle. 

This paper is organized as follows. 
In section 2 we briefly review the Teukolsky formalism. 
In section 3 we discuss the equations of motion and 
the energy momentum tensor of a spinning particle. 
In section 4 we solve the equations of motion 
to the linear order of the amplitude of spin. 
There we obtain a family of `circular' orbits which have vanishing 
radial velocity and stay close to the equatorial plane. 
In section 5 we evaluate gravitational waves from
the spinning particle in circular orbits 
and give the formula for the energy loss rate to 2.5PN order. 
In section 6 we consider the problem of the radiation reaction. 
There we show that the assumption that the orbit remains circular
under radiation reaction is consistent with the energy and angular
momentum loss rates in the linear order of spin.
Then we evaluate the rate of change in the orbital frequency 
under this assumption. 
In section 7, brief summary and discussion are in order. 

We use the units $G=c=1$ and the metric signature $(-,+,+,+)$. 
The round (square) brackets on the indices denote
the (anti-) symmetrization, e.g.,
$$
  \Phi_{(\mu\nu)}={1\over 2}\left(\Phi_{\mu\nu}+\Phi_{\nu\mu}\right),
\quad
  \Phi_{[\mu\nu]}={1\over 2}\left(\Phi_{\mu\nu}-\Phi_{\nu\mu}\right).
$$

\section{Teukolsky Formalism}

In this section we briefly review the Teukolsky formalism. 
For details, see e.g., Ref.\cite{suppl} and references cited therein.
In the Teukolsky formalism, the 
wave form and the energy flux of gravitational waves are calculated 
from the fourth Newman-Penrose quantity\cite{New}, which is 
expanded as 
\begin{equation}
\psi_4=(r-ia\cos\theta)^{-4}\int d\omega e^{-i\omega t}\sum_{\ell,m}
e^{im\varphi}{_{-2}S^{a\omega}_{\ell m}(\theta) \over \sqrt{2\pi}}
R_{\ell m\omega}(r).
\end{equation}
Here, $_{-2}S^{a\omega}_{\ell m}(\theta)$ is the 
spin weighted spheroidal harmonics 
normalized by
\begin{equation}
 \int_0^{\pi}\vert {}_{-2}S^{a\omega}_{\ell m}(\theta)\vert^2 
 \sin\theta d\theta =1, 
\end{equation}
and its eigen value is $\lambda$. 
Then $R_{\ell m\omega}$ obeys the Teukolsky equation 
\begin{equation}
\Delta^2 {d \over dr}\Bigl({1 \over \Delta}{dR_{\ell m\omega} \over dr}\Bigr)
-V(r)R_{\ell m\omega}=T_{\ell m\omega}(r),
\end{equation}
and 
\begin{equation}
V(r)=-{K^2+4i(r-M)K \over \Delta}+8i\omega r+\lambda,
\end{equation}
where $\Delta=r^2-2Mr+a^2$ and $K=(r^2+a^2)\omega-ma$. 
The source term $T_{\ell m\omega}(r)$ is constructed
from the energy momentum tensor of the particle, and its
explicit form is given later. 

The solution of the Teukolsky equation at infinity 
($r\rightarrow\infty$) is expressed as
\begin{equation}
R_{\ell m\omega}(r)
 \rightarrow{r^3e^{i\omega r^*} \over 2i\omega B^{in}_{\ell m\omega}}
\int^{\infty}_{r_+}dr'{T_{\ell m\omega}(r') R_{\ell m\omega}^{in}(r') 
\over\Delta^{2}(r')}
\equiv \tilde Z_{\ell m\omega}r^3e^{i\omega r^*},
\label{Rinfty}
\end{equation}
where $r_+=M+\sqrt{M^2-a^2}$ denotes the radius of the event horizon
and $R_{\ell m\omega}^{in}$ is the homogeneous solution
which satisfies the ingoing-wave boundary condition at horizon,
\begin{equation}
R_{\ell m\omega}^{in} \rightarrow
\left\{\begin{array}{lr}
\displaystyle D_{\ell m\omega}\Delta^2  e^{-ikr^*},& \quad
r^*\rightarrow -\infty, \\
\displaystyle r^3 B_{\ell m\omega}^{out}e^{i\omega r^*}+
r^{-1}B_{\ell m\omega}^{in} e^{-i\omega r^*}, & \quad
r^*\rightarrow +\infty, \\
\end{array}\right.
\label{Rin}
\end{equation}
where $k=\omega-ma/2Mr_+$ and
$r^*$ is the tortoise coordinate defined by 
\begin{equation}
{dr^* \over dr}={r^2+a^2 \over \Delta}.
\end{equation}
For definiteness, we fix the integration constant such that
$r^*$ is given explicitly by
\begin{eqnarray}
r^{*}&=&\int {dr^*\over dr}dr \nonumber \\
&= & r+{2Mr_{+}\over {r_{+}-r_{-}}}\ln{{r-r_{+}}\over 2M}
-{2Mr_{-}\over {r_{+}-r_{-}}}\ln{{r-r_{-}}\over 2M},
\label{rst}
\end{eqnarray}
where $r_{\pm}=M\pm \sqrt{M^2-a^2}$.

Thus in order to calculate gravitational waves emitted to infinity from
a particle in circular orbits,
we need to know the explicit form of the source term $T_{\ell m\omega}(r)$,
which has support only at $r=r_0$ where $r_0$ is the orbital radius
in the Boyer-Lindquist coordinate,
the ingoing-wave Teukolsky function $R^{in}_{\ell m\omega}(r)$ at $r=r_0$,
and its incident amplitude $B^{in}_{\ell m\omega}$ at infinity.
We consider the expansion of these 
quantities in terms of a small parameter, 
$v^2 \equiv M/r_0$. Note that $v$ is approximately 
equal to the orbital velocity, but not strictly equal to it 
in the case of $a \ne 0$ or $S\ne 0$.
A systematic expansion method to calculate these necessary quantities
has been developed in refs.\cite{SSTT,SSTT2}, by considering the
Sasaki-Nakamura equation first and then transforming the result to
Teukolsky equation.

In addition to these, we need to expand the
spheroidal harmonics and their eigenvalues in powers of $a\omega$.
Since $\omega=O(\Omega)$ where $\Omega$ is the orbital angular
velocity of the particle, we have $a\omega=O(M\omega)=O(v^3)$. 
Thus the expansion in powers of $a\omega$ is also a part of the
post-Newtonian expansion.
Note also that the spin parameter of the black hole $a$ does not 
have to be small but can be of order $M$.

The expressions of these quantities required to calculate the 
energy loss rate up to 2.5PN order are already obtained
in ref. \cite{SSTT}. 
We digest the results omitting all the derivations. 
The homogeneous solutions of the Teukolsky equation with 
the ingoing boundary condition for $\ell=2,3,4$ are 
\begin{eqnarray}
\omega R^{in}_{2m\omega}&= &
{{{z^4}}\over {30}} + {i\over {45}}\,{z^5} 
- {{11\,{z^6}}\over {1260}} -   {i\over {420}}\,{z^7} 
+ {{23\,{z^8}}\over {45360}} + {i\over {11340}}\,{z^9}
\nonumber \\
&&+ \epsilon\Bigl(
{{-{z^3}}\over {15}} - {i\over {60}}\,m\,q\,{z^3} 
- {i\over {60}}\,{z^4} + {{m\,q\,{z^4}}\over {45}}
- {{41\,{z^5}}\over {3780}} +   {{277\,i}\over {22680}}\,m\,q\,{z^5} 
- {{31\,i}\over {3780}}\,{z^6} -  {{7\,m\,q\,{z^6}}\over {1620}}
\Bigr)
\nonumber \\
& & + \epsilon^2\Bigl(
{{{z^2}}\over {30}} + {i\over {40}}\,m\,q\,{z^2} + 
  {{{q^2}\,{z^2}}\over {60}} - {{{m^2}\,{q^2}\,{z^2}}\over {240}} 
- {i\over {60}}\,{z^3} - {{m\,q\,{z^3}}\over {30}} 
+ {{i}\over {90}}\,{q^2}\,{z^3} 
- {i\over {120}}\,{m^2}\,{q^2}\,{z^3}\Bigr),
\label{elltwo} \\\\
\omega R^{in}_{3m\omega}&=&
{{{z^5}}\over {630}} + {i\over {1260}}\,{z^6} - {{{z^7}}\over {3780}} 
- {i\over {16200}}\,{z^8} +\epsilon\Bigl(
{{-{z^4}}\over {252}} - {i\over {1890}}\,m\,q\,{z^4} 
- {i\over {756}}\,{z^5} + {{11\,m\,q\,{z^5}}\over {22680}}
\Bigr),
\\\\
\omega R^{in}_{4m\omega}&= &{z^6\over 11340}+{iz^7\over 28350}.
\end{eqnarray}
where $\epsilon:=2M\omega$, $z:=\omega r$ and $q:=a/M$.  
The incident amplitudes are 
\begin{eqnarray} 
B^{in}_{2m\omega} & = &
{i \over 8\omega^2}\Bigl\{ 1-\epsilon{\pi\over 2}
+i\epsilon \Bigl({5 \over 3}-\gamma -\log 2\Bigr)+
{m q \over 18}\epsilon +O(\epsilon^2)\Bigr\} ,
\\
B^{in}_{3m\omega} & = & 
-{1 \over 8\omega^2}\Bigl\{ 1-\epsilon{\pi\over 2}
+i\epsilon \Bigl({13 \over 6}-\gamma -\log 2\Bigr)+
{m q \over 72}\epsilon +O(\epsilon^2) \Bigr\},
\\
B^{in}_{4m\omega}& = &-{i \over 8\omega^2}\Bigl\{1+O(\epsilon)\Bigr\}.
\end{eqnarray}

\section{Spinning Particle}

To give the source term of the Teukolsky equation, 
we need to solve the motion of the spinning particle 
and also to give the expression of the energy momentum tensor. 
In this section we give the necessary expressions, following
refs.\cite{dixon,wald,ours}.

Neglecting the effect of the higher multipole moments, 
the equations of motion of a spinning particle are given by
\begin{eqnarray}
{D\over d\tau}p^\mu(\tau)  & = & 
-{1\over 2} R^\mu_{~\nu\rho\sigma}(z(\tau))
 v^\nu(\tau) S^{\rho\sigma}(\tau), 
\nonumber \\
{D\over d\tau}S^{\mu\nu}(\tau) & = & 
2 p^{[\mu}(\tau)v^{\nu]}(\tau), 
\label{eqofmot}
\end{eqnarray}
where $v^{\mu}(\tau)=dz^{\mu}(\tau)/d\tau$, $\tau$ is a parameter
which is not necessarily the proper time of the particle,
and, as we will see later, the vector $p^{\mu}(\tau)$ and the 
antisymmetric tensor $S^{\mu\nu}(\tau)$ represent 
the linear and spin angular momenta of the particle, 
respectively. Here $D/d\tau$ denotes the covariant derivative along
the particle trajectory.

We do not have the evolution equation for $v^{\mu}(\tau)$ yet.
In order to determine $v^\mu(\tau)$,
we need to impose a supplementary condition which determines the center 
of mass of the particle\cite{dixon},
\begin{equation}
S^{\mu\nu}(\tau)p_{\nu}(\tau)=0. 
\label{centmass}
\end{equation}
Then one can show that $p_\mu p^\mu=const.$ and
$S_{\mu\nu}S^{\mu\nu}=const.$ along the particle trajectory\cite{wald}.
Therefore we may set
\begin{eqnarray}
p^\mu&=&\mu u^\mu\,,\quad u_\mu u^\mu=-1,
\nonumber\\
S^{\mu\nu}&=&\epsilon^{\mu\nu}{}_{\rho\sigma}p^\rho S^\sigma\,,
\quad p_\mu S^\mu=0,
\nonumber\\
S^2&=&S_\mu S^\mu={1\over2\mu^2}S_{\mu\nu}S^{\mu\nu},
\end{eqnarray}
where $\mu$ is the mass of the particle,
 $u^\mu$ is the specific linear momentum,
and $S^\mu$ is the specific spin vector
with $S$ its magnitude. Note that if we use $S^\mu$ instead of
$S^{\mu\nu}$ in the equations of motion, the center of mass condition
(\ref{centmass}) will be replaced by the condition
$p_\mu S^\mu=0$ (see section 4).

Since the above equations of motion are invariant under 
reparametrization of the orbital parameter $\tau$,
 we can fix $\tau$ to satisfy  
\begin{equation}
 u^\mu(\tau)v_\mu(\tau)=-1. 
\label{normvel}
\end{equation}
Then, from Eqs.~(\ref{eqofmot}, (\ref{centmass}) and (\ref{normvel}), 
$v^\mu(\tau)$ is given as\cite{dixon}
\begin{equation}
v^\mu(\tau)-u^\mu(\tau)
={1\over 2}
\left(\mu^2+{1\over 4}R_{\chi\xi\zeta\eta}(z(\tau))S^{\chi\xi}(\tau)
S^{\zeta\eta}(\tau)\right)^{-1} S^{\mu\nu}(\tau)
R_{\nu\rho\sigma\kappa}(\tau)u^\rho(\tau)S^{\sigma\kappa}(\tau). 
\label{vprelation}
\end{equation}
With this equation, the equations of motion (\ref{eqofmot}) 
completely determine the evolution of the orbit and the spin.

As for the energy momentum tensor, Dixon\cite{dixon} gives it in 
terms of the Dirac delta-function on the tangent space at 
$x^\mu=z^\mu(\tau)$. For later convenience, in this paper
 we use an equivalent but
alternative form of the energy momentum tensor, given in
terms of the Dirac delta-function on the coordinate space\cite{ours}:
\begin{eqnarray}
T^{\alpha\beta}(x) = &&\int d\tau\Biggl\{ 
p^{(\alpha}(x,\tau)v^{\beta)}(x,\tau)
{\delta^{(4)}\left(x-z(\tau)\right)\over\sqrt{-g}}
\nonumber \\
&&\quad -\nabla_{\gamma}\left(S^{\gamma(\alpha}(x,\tau)v^{\beta)}(x,\tau)
{\delta^{(4)}
\left(x-z(\tau)\right)\over\sqrt{-g}}\right)\Biggr\}, 
\label{dixontensor}
\end{eqnarray}
where $v^{\alpha}(x,\tau)$, $p^{\alpha}(x,\tau)$ 
and $S^{\alpha\beta}(x,\tau)$ are bi-tensors 
which are spacetime extensions of $v^{\mu}(\tau)$, $p^{\mu}(\tau)$
and $S^{\mu\nu}(\tau)$ which are defined only along the world line, 
$x^\mu=z^\mu(\tau)$.\footnote{In the rest of this section, we use 
$\mu,\nu,\sigma,\cdots$ as the tensor indices associated 
with the world line $z(\tau)$ and $\alpha,\beta,\gamma,\cdots$ as 
those with a field point $x$, and suppress the coordinate indices
of $z(\tau)$ and $x$ for notational simplicity.}
To define $v^{\alpha}(x,z(\tau))$, $p^{\alpha}(x,z(\tau))$
and $S^{\alpha\beta}(x,z(\tau))$ 
we introduce a bi-tensor $\bar g^{\alpha}_{~\mu}(x,z)$ which satisfies 
\begin{eqnarray}
&&\lim_{x\rightarrow z}\bar g^{\alpha}_{~\mu}(x,z(\tau))
=\delta^{\alpha}_{~\mu}\,, 
\nonumber \\
&&\lim_{x\rightarrow z}\nabla_{\beta}\bar g^{\alpha}_{~\mu}(x,z(\tau))=0.  
\end{eqnarray}
For the present purpose, further specification of 
$\bar g^{\alpha}_{~\mu}(x,z)$ is not necessary. 
Using this bi-tensor $\bar g^{\alpha}_{~\mu}(x,z)$, we define 
$p^\alpha(x,\tau)$, $v^\alpha(x,\tau)$ and $S^{\alpha\beta}(x,\tau)$ as 
\begin{eqnarray}
p^\alpha(x,\tau)&=&\bar g^{\alpha}_{~\mu}
      \bigl(x,z(\tau)\bigr)p^\mu (\tau), 
\nonumber \\
v^\alpha(x,\tau)&=&\bar g^{\alpha}_{~\mu}
      \bigl(x,z(\tau)\bigr)v^\mu (\tau), 
\nonumber \\
S^{\alpha\beta}(x,\tau)&=&\bar g^{\alpha}_{~\mu}\bigl(x,z(\tau)\bigr)
\bar g^{\beta}_{~\nu}
      \bigl(x,z(\tau)\bigr)S^{\mu\nu}(\tau). 
\end{eqnarray}

It is easy to see that the divergence free condition of 
this energy momentum tensor gives the equations of motion
(\ref{eqofmot}). 
Noting the relations,
\begin{eqnarray}
\nabla_\beta\bar g^{\alpha}_{~\mu}(x,z(\tau)) 
\delta^{(4)}(x,z(\tau)) & = & 0, 
\nonumber \\
v^{\alpha}(x)\nabla_{\alpha}\left({\delta^{(4)}\bigl(x,z(\tau)\bigr)\over
\sqrt{-g}}\right) & = & 
  -{d\over d\tau}\left({\delta^{(4)}\bigl(x,z(\tau)\bigr)
\over\sqrt{-g}}\right), 
\end{eqnarray}
the divergence of Eq.~(\ref{dixontensor}) becomes
\begin{eqnarray}
\nabla_{\beta}T^{\alpha\beta}(x)
&=&\int d\tau \bar g^{\alpha}_{~\mu}\bigl(x,z(\tau)\bigr) 
{ \delta^{(4)} \bigl(x-z(\tau)\bigr) \over \sqrt{-g} }
\left( {d\over d\tau}p^{\mu}(\tau)
+{1\over 2}R^{\mu}_{~\nu\sigma\kappa}\bigl(z(\tau)\bigr)
v^{\nu}(\tau) S^{\sigma\kappa}(\tau) \right)
\nonumber \\ 
&&~+{1\over 2} \int d\tau \nabla_\beta\left(
\bar g^{\alpha}_{~\mu}\bigl(x,z(\tau)\bigr) 
\bar g^{\beta}_{~\nu}\bigl(x,z(\tau)\bigr) 
{ \delta^{(4)} \bigl(x-z(\tau)\bigr) \over \sqrt{-g} } \right)
\left( {d\over d\tau}S^{\mu\nu}(\tau)
-2p^{[\mu}(\tau)v^{\nu]}(\tau) \right). 
\end{eqnarray}
Since the first and second terms in the right-hand side must 
vanish separately, 
we obtain the equations of motion~(\ref{eqofmot}).

In order to clarify the meaning of $p^{\mu}$ and $S^{\mu\nu}$, 
we consider the volume integral of this energy momentum tensor such as 
$\displaystyle \int_{\Sigma(\tau_0)} \bar g^{~\mu}_{\alpha}
         T^{\alpha\beta} d\Sigma_{\beta}$, 
where we take the surface ${\Sigma(\tau_0)}$ to be
perpendicular to $u^{\alpha}(\tau_0)$. 
It is convenient to introduce a scalar function $\tau(x)$ 
which determines the surface ${\Sigma(\tau_0)}$ by the equation
$\tau(x)=\tau_0$, and 
${\partial\tau/\partial x^{\beta}} =-u_{\beta}$ at $x=z(\tau_0)$. 
Then we have
\begin{eqnarray}
\int_{\Sigma(\tau_0)} \bar g^{~\mu}_{\alpha} 
         T^{\alpha\beta} d\Sigma_{\beta}
& = &
\int d^4 x \sqrt{-g} {\partial\tau\over \partial x^{\beta}}
 \delta(\tau(x)-\tau_0) \bar g^{~\mu}_{\alpha}T^{\alpha\beta}(x)
\nonumber \\
&=&\int d\tau' \Biggl\{\delta(\tau'-\tau_0) 
 \left[p^{\mu}+p^{[\mu} v^{\nu]}u_{\nu}
 -{1\over 2} {D u_{\nu}\over d\tau} S^{\nu\mu}\right] \Biggr\}
\nonumber \\
& = &p^{\mu}(\tau_0).
\end{eqnarray}
where we used the center of mass condition and the equation of motion
for $S^{\mu\nu}$. We clearly see $p^\mu$ indeed represents the linear
momentum of the particle.

In order to clarify the meaning of $S^{\mu\nu}$, 
following Dixon \cite{dixon}, we introduce the relative position vector
\begin{equation}
 X^{\mu}:= -g^{\mu\nu}\partial_\nu\sigma(x,z), 
\end{equation}
where $\sigma(x,z)$ is the geodetic interval between $z$ and $x$
defined by using the parametric form of a 
geodesic $y(u)$ joining $z=y(0)$ and $x=y(1)$ as 
\begin{equation}
\sigma(x,z):={1\over 2}\int_0^1 g_{\alpha\beta}
       {dy^{\alpha}\over du}{dy^{\beta}\over du} du.
\end{equation}
Then noting the relations 
\begin{equation}
\lim_{x\rightarrow z}X^{\mu}=0, \quad
\lim_{x\rightarrow z}X^{\mu}_{~,\beta}=\delta^{\mu}_{\beta},
\end{equation}
it is easy to see that 
\begin{equation}
S^{\mu\nu}=2\int_{\Sigma_{\tau_0}}
      X^{[\mu}\bar g^{~\nu]}_{\alpha} T^{\alpha\beta} d\Sigma_{\beta}.
\end{equation}
Now that the meaning of $S^{\mu\nu}$ is manifest. 
>From the above equation, it is also easy to see
that the center of mass condition (\ref{centmass}) is the 
generalization of the Newtonian counter part,
\begin{equation}
 \int d^3 x \rho(x) x^i =0,
\end{equation}
where $\rho$ is the matter density. 

Before closing this section, we mention several conserved quantities
of the present system. We have already noted that $p_\mu p^\mu=-\mu^2$
and $S_\mu S^\mu=S^2$ are 
constant along the particle trajectory on an arbitrary spacetime.
There will be an additional conserved
quantity if the spacetime admits a Killing vector field $\xi_{\mu}$, 
\begin{equation}
 \xi_{(\mu;\nu)}=0.
\end{equation}
Namely, the quantity
\begin{equation}
 Q_\xi:=p^{\mu}\xi_{\mu}-{1\over 2}S^{\mu\nu}\xi_{\mu;\nu}, 
\label{cons}
\end{equation}
is conserved along the particle trajectory\cite{dixon}. 
It is easy to verify that $Q_\xi$ is conserved by directly
using the equations of motion.

\section{Circular orbits}

Let us consider ``circular'' orbits in the Kerr spacetime 
with a fixed Boyer-Lindquist radial coordinate, $r=r_0$.
We consider a class of orbits that would stay on the equatorial plane
if the particle were spinless. Hence we
assume that $\tilde \theta :=\theta -\pi/2 \sim
O(S/M)\ll 1$.
Under this assumption, we write down the equations of motion
and solve them up to the linear order in $S$.
In appendix we give a further analysis in the case in which the 
spin vector (see below) is parallel or anti-parallel to the rotation 
axis of the black hole. 

In order to find a solution representing a circular orbit, 
it is convenient to introduce the  tetrad frame defined by 
\begin{eqnarray}
e^{0}_{~\mu}   &= &\Bigl(\sqrt{{\Delta \over \Sigma}}, ~0, ~0, 
-a\sin^2\theta\sqrt{{\Delta \over \Sigma}}\Bigr),
\nonumber \\
e^{1}_{~\mu}   &= &\Bigl(0, ~\sqrt{{\Sigma \over \Delta}}, ~0, ~0 \Bigr),
\nonumber \\
e^{2}_{~\mu}&= & \Bigl(0,~0, ~\sqrt{\Sigma}, ~0 \Bigr),
\nonumber \\
e^{3}_{~\mu}&= &\Bigl( -{a \over \sqrt{\Sigma} }\sin\theta 
,~0 ,~0 ,~{ r^2+a^2 \over \sqrt{\Sigma}} \sin\theta \Bigr),
\end{eqnarray}
where 
$\Sigma=r^2+a^2\cos^2\theta$, and
$e^a_{~\mu}=(e^a_{~t},e^a_{~r},e^a_{~\theta},e^a_{~\varphi})$ 
for $a=0\sim 3$. Hereafter,
we use the Latin letters to denote the tetrad indices. 

For convenience, we introduce $\omega_1 \sim \omega_6$ to represent 
the tetrad components of the spin coefficients near the equatorial
plane:
\begin{eqnarray}
 \omega_{01}^{~~0} & = &\omega_{00}^{~~1}=
 \omega_1+O(\tilde \theta^2),\quad 
 \omega_1=
 {a^2 -Mr\over r^2 \Delta^{1/2}},
\nonumber \\
 \omega_{31}^{~~0} & = &\omega_{30}^{~~1}=
 \omega_{13}^{~~0} = \omega_{10}^{~~3}=
 \omega_{03}^{~~1} = -\omega_{01}^{~~3}=
 \omega_2+O(\tilde \theta^2),\quad
 \omega_2
 :={a\over r^{2}}
\nonumber \\
 \omega_{22}^{~~1} & = &-\omega_{21}^{~~2}=
 \omega_{33}^{~~1} = -\omega_{31}^{~~3}=
 \omega_3+O(\tilde \theta^2),\quad
 \omega_3
 :={\Delta^{1/2}\over r^2},
\nonumber \\
 \omega_{02}^{~~0} & = &\omega_{00}^{~~2}=
 \omega_{12}^{~~1} = -\omega_{11}^{~~2}=\tilde \theta~ \omega_4
 +O(\tilde \theta^2),\quad
 \omega_4
 :=-{a^2\over r^3},
\nonumber \\
 \omega_{32}^{~~0} & = &\omega_{30}^{~~2}=
 -\omega_{23}^{~~0} = -\omega_{20}^{~~3}=
 \omega_{03}^{~~2} = -\omega_{02}^{~~3}=
 \tilde \theta~ \omega_5+O(\tilde \theta^2),\quad  
 \omega_5
 :=-{a\Delta^{1/2}\over r^{3}}, 
\nonumber \\
 \omega_{33}^{~~2} & = &-\omega_{32}^{~~3}=
 \tilde \theta~ \omega_6+O(\tilde \theta^2),\quad
 \omega_6 
 := -{(r^2+a^2)\over r^{3}},
\end{eqnarray}
where $\omega_{ab}^{~~c}=e^{~\mu}_a e^{~\nu}_b e^c{}_{\nu;\mu}$. 
Since the following relation holds for an arbitrary vector $f^{\mu}$, 
$$
 e^a_{~\mu} {D \over d\tau}f^{\mu}=
 {d \over d\tau}f^{a}-\omega_{bc}^{~~a}v^b f^c, 
$$
the tetrad components of $Df^\mu/d\tau$ along a circular orbit
are given explicitly as 
\begin{eqnarray}
 e^0_{~\mu} 
{D\over d\tau}f^{\mu}&=&\dot f^{0} -\left( A f^1 +\tilde\theta C f^2\right)
+O(\tilde \theta^2),
\nonumber \\
e^1_{~\mu} 
 {D\over d\tau}f^{\mu}&=&\dot f^{1} -\left( A f^0 +B f^3 +E f^2\right)
+O(\tilde \theta^2),
\nonumber \\
e^2_{~\mu} 
 {D\over d\tau}f^{\mu}&=&\dot f^{2} -\left( 
         \tilde\theta C f^0 
+\tilde\theta D f^3-Ef^1\right)+O(\tilde \theta^2),
\nonumber \\
e^3_{~\mu} 
 {D\over d\tau}f^{\mu}&=&\dot f^{3} -\left( -B f^1 -\tilde\theta D f^2\right)
+O(\tilde \theta^2), 
\end{eqnarray}
where $A$, $B$, $C$, $D$ and $E$ are defined by\footnote{The 
symbols $A\sim E$ used here to define the auxiliary variable
are applicable only in this section, and not to be
confused with quantities defined with the same symbols. such as
$E$ for energy, in the later sections.}
\begin{eqnarray}
  A&:=&\omega_1 v^0 +\omega_2 v^3,
\nonumber \\
  B&:=&\omega_2 v^0 +\omega_3 v^3, 
\nonumber \\
  C&:=&\omega_4 v^0 +\omega_5 v^3,
\nonumber \\
  D&:=&\omega_5 v^0 +\omega_6 v^3,
\nonumber \\
  E&:=&\omega_3 v^2, 
\end{eqnarray}
and we have assumed that $v^1=0$ and $v^2=O(\tilde \theta)$. 

Now we rewrite the equations of motion changing the spin variable.  
We replace the spin tensor with the unit spin 
vector, $\zeta^a$, which is defined by
\begin{equation}
  \zeta^a:={S^a\over S}=-{1\over 2\mu S}\epsilon^a_{~bcd} u^b S^{cd},
\end{equation}
or equivalently by
\begin{equation}
 S^{ab}= \mu S \epsilon^{ab}_{~~cd} u^c \zeta^{d},
\end{equation}
where $\epsilon_{abcd}$ is the completely antisymmetric symbol with
the convention of $\epsilon_{0123}=1$.
As noted in the previous section, if we use the spin vector as an 
independent variable, 
the center of mass condition is automatically satisfied, while 
it becomes necessary to impose another supplementary condition 
\begin{equation}
\zeta^a u_a=0.
\end{equation} 
Then the equations of motion reduce to 
\begin{eqnarray}
 {d u^a\over d\tau} & = & \omega_{bc}^{~~a}v^b u^c -SR^a,
\nonumber \\
 {d \zeta^a\over d\tau} & = & \omega_{bc}^{~~a}v^b \zeta^c 
                       -Su^a \zeta^b R_{b},
\end{eqnarray}
where 
\begin{equation}
  R^a:=R^{*a}{}_{bcd}v^b u^c \zeta^d
   ={1\over 2\mu S}R^a{}_{bcd} v^b S^{cd},
\end{equation}
and $ R^{*}_{abcd}={1\over 2}R_{abef}\epsilon^{ef}{}_{cd}$ 
is the right dual of the Riemann tensor.
It will be convenient to write explicitely the tetrad components of 
$ R^{*}_{~abcd}$.  
Since we only need $R^{*}_{abcd}$ at $O(\tilde \theta^0)$, 
the non-vanishing components of $ R^{*}_{~abcd}$ are given by
\begin{equation}
 -{1\over 2}R^{*}_{0123} =- R^{*}_{0213}=
 R^{*}_{0312}= R^{*}_{1203}=
 -R^{*}_{1302}= -{1\over 2}R^{*}_{2301}=
 -{M\over r^3}+O(\tilde \theta^2).
\end{equation}
Although we do not need them, we note that the following components are 
not identically zero but are of $O(\tilde \theta)$. 
$$
R^{*}_{1212},~ R^{*}_{1313},~ R^{*}_{1010},~ R^{*}_{2323},~
R^{*}_{2020}, ~{\rm and~} R^{*}_{3030}. 
$$

\subsection{Lowest Order in $S$}

We first solve the equations of motion for a circular orbit at $r=r_0$
at the lowest order in $S$. 
For notational simplicity, we omit the suffix $0$ of $r_0$
in the following.
For the class of orbits we have assumed, we have $v^1=0$ and
$v^2=O(\tilde \theta)$.Then the non-trivial equations are 
\begin{eqnarray}
 {d\over d\tau}v^1&=& A v^0 + B v^3 =0,
\label{v1} \\
\nonumber\\
 {d\over d\tau} \zeta^2 &= &0,
\quad
 {d\over d\tau} \left(
  \begin{array}{c}
   \zeta^0 \\ \zeta^1 \\ \zeta^3
  \end{array}\right)
 = \left(
  \begin{array}{ccc}
    0 & A & 0 \\
    A & 0 & B \\
    0 & -B & 0 \\
  \end{array}\right)
   \left(\begin{array}{c}
   \zeta^0 \\ \zeta^1 \\ \zeta^3
  \end{array}\right).
\label{spineq0}
\end{eqnarray}
The equation (\ref{v1}) determines the rotation velocity of the 
orbital motion. By setting $x:=v^3/v^0$, we obtain the equation
\begin{equation}
 \omega_1 +2\omega_2 x +\omega_3 x^2=0,  
\end{equation}
which is solved to give
\begin{equation}
 x={\pm\sqrt{Mr}-a\over \sqrt{\Delta}}. 
\end{equation}
The upper (lower) sign corresponds to the case that 
$v^3$ is positive (negative).
Then, with the aid of the normalization condition of four momentum, 
$v^{\mu} v_{\mu}=-1+O(S^2)$, we find
\begin{equation}
 v^0={1\over\sqrt{1-x^2}},\quad  v^3={x\over\sqrt{1-x^2}}.
\label{xtov}
\end{equation}
Note that, in this case, the orbital angular frequency $\Omega$ 
is given by a well known formula,
\begin{equation}
\Omega={\pm \sqrt{M} \over r^{3/2} \pm \sqrt{M}a}\,. 
\end{equation}
On the other hand, the equations of spin (\ref{spineq0}) are
solved to give
\begin{equation}
 \zeta^2=-\zeta_{\perp}\,,
\qquad
\left(\begin{array}{c}
   \zeta^0 \\ \zeta^1 \\ \zeta^3
  \end{array}\right)
 =\zeta_\parallel\left(
  \begin{array}{c}
   \alpha\sin(\phi+c_1)+\beta c_2\\
    \cos(\phi+c_1)  \\
   -\beta\sin(\phi+c_1)-\alpha c_2
\end{array}\right).
\end{equation}
where $\zeta_{\perp}$, $\zeta_{\parallel}$, $c_1$ and $c_2$ are
constants, and
\begin{eqnarray}
\alpha&=&{A\over\sqrt{B^2-A^2}}=\mp v^3,
\quad \beta={B\over\sqrt{B^2-A^2}}=\pm v^0, 
\nonumber \\
 \phi & = & \Omega_p \tau,\quad \Omega_p=\sqrt{B^2-A^2} =\sqrt{{M \over
r^3}}\,.
\end{eqnarray}
The supplementary condition $v^a \zeta_a=0$ requires that $c_2=0$. The
condition $\zeta_a\zeta^a=1$ implies
$\zeta_\perp^2+\zeta_\parallel^2=1$.  Further since the origin of the
time $\tau$ can be chosen arbitrarily, we set $c_1=0$. Thus, we
obtain
\begin{equation}
\zeta^2=-\zeta_\perp\,,\qquad\left(
  \begin{array}{c}
 \zeta^0 \\ 
 \zeta^1 \\ 
 \zeta^3 \end{array}
\right)
=\zeta_{\parallel}
\left( \begin{array}{r} 
   \alpha\sin\phi \\ 
        \cos\phi \\
 -\beta\sin\phi
\end{array}\right).
\end{equation}
Here, we should note that $\Omega_p \not= \Omega$ in general if $a
\not= 0$ or $S \not=0$ (see below).

\subsection{Next Order}

Having obtained the leading order solution with respect to $S$, 
we now turn to the equations of motion up to the linear order in $S$. 
We assume that the spin vector components are expressed in 
the same form as were in the leading order but consider 
corrections to the coefficients $\alpha, \beta$ and $\Omega_p$ 
of order $S$. 
As long as we are working only up to the linear order in $S$, 
Eq.~(\ref{vprelation}) tells us that $v^a$ can be identified 
with $u^a$.
In order to write down the equations of motion up to the linear 
order in $S$, we need the explicit form of $R^a$, which can be evaluated 
by using the knowledge of the lowest order solution, as 
\begin{eqnarray}
 R^0 & = & R^3 =O(\tilde \theta), 
 \nonumber \\
 R^1 & = & 3{M\over r^3} v^0 v^3 \zeta^2+O(\tilde \theta), 
\nonumber \\
 R^2 & = & 3{M\over r^3} v^0 v^3 \zeta^1+O(\tilde \theta).
\end{eqnarray}

First we consider the orbital equations of motion.
With the assumption that $v^1=0$ and $v^2=O(\tilde \theta)$, 
the non-trivial equations of the orbital motion are
\begin{eqnarray}
 \dot v^1 & = & A v^0+B v^3- SR^1=0,
 \label{v1sec} \\
 \dot v^2 & = & (C v^0+D v^3)\tilde \theta - SR^2. 
 \label{v2sec} 
\end{eqnarray}
The first equation gives the rotation velocity as before,
while the second equation determines the motion in the 
$\theta$-direction.

Again using the variable $x=v^3/v^0$, Eq.~(\ref{v1sec}) is rewritten as 
\begin{equation}
\omega_1+2 \omega_2 x+\omega_3 x^2+3 {S_{\perp}M\over r^3} x=0, 
\end{equation} 
where $S_{\perp}:=S \zeta_{\perp}$. 
The solution of this equation is 
\begin{equation}
 x=\left({\pm\sqrt{Mr}-a\over\sqrt{\Delta}}\right)
         \left(1\mp {3S_{\perp}\sqrt{M}\over 2 r^{3/2}}\right)+O(S^2).
\end{equation}
Using the relations (\ref{xtov}), it immediately gives $v^0$ and $v^3$.
{}From the definition of the tetrad, we have the following relations,  
\begin{eqnarray}
 v^0 = \sqrt{\Delta\over \Sigma}\left[{dt\over d\tau}
    -a\sin^2\theta{d\varphi\over d\tau}\right],
 \nonumber \\
 v^3 = {\sin\theta\over \sqrt{\Sigma}}\left[-a{dt\over d\tau}
    +(r^2+a^2){d\varphi\over d\tau}\right]. 
\label{v0v3}
\end{eqnarray}
Thus, the orbital angular velocity observed at infinity is calculated 
to be
\begin{equation}
 \Omega:={d\varphi\over dt}
={a+x\sqrt{\Delta} \over r^2+a^2+ax\sqrt{\Delta}}+O(\tilde \theta^2)
    =\pm{\sqrt{M}\over r^{3/2}\pm a\sqrt{M}}
  \left[1-{3 S_{\perp}\over 2}{\pm\sqrt{Mr}
          -a\over r^2\pm a\sqrt{Mr}}\right]+O(\tilde \theta^2).
  \label{Omegaori}
\end{equation}

In order to solve the second equation (\ref{v2sec}),
we note that $v^2=\sqrt{\Sigma}\dot\theta\simeq r \dot {\tilde\theta}$ and
\begin{equation}
 Cv^0+Dv^3=-{M\over r^2}{1+2 x^2\over 1-x^2} +O(S).
\end{equation}
Then we find that Eq.~(\ref{v2sec}) reduces to 
\begin{equation}
 r \ddot {\tilde\theta}= -{M\over r^2}{1+2 x^2\over 1-x^2}{\tilde\theta}
            -3 {S_{\parallel}M\over r^3}{x\over 1-x^2}\cos\phi\,, 
\label{tildetheta}
\end{equation}
where $S_{\parallel}=S \zeta_{\parallel}$.
This equation can be solved easily by setting $\tilde\theta
=\theta_0 \cos\phi$. 
Recalling that $\Omega_p^2=M/r^3+O(S)$, we obtain
\begin{equation}
 \theta_0= -{S_{\parallel}\over rx}.
\end{equation}
Thus we see that the orbit will remain in the equatorial plane if
$S_\parallel=0$, but deviates from it if $S_\parallel\neq0$.
We note that
there exists a degree of freedom to add a homogeneous solution of 
Eq.~(\ref{tildetheta}),
whose frequency, $\Omega_{\theta}=\displaystyle\sqrt{{M\over r^3}{
1+2x^2\over 1-x^2}}$, is different from $\Omega_p$ 
and which corresponds to giving a small inclination angle to the orbit,
indifferent to the spin.
Here, for simplicity, we only consider the case when 
this homogeneous solution to $\tilde\theta$ is zero.
Schematically speaking, the orbits under consideration
are those with the total angular momentum {\boldmath$J$} being
parallel to the $z$-direction, which is sum of the orbital and spin 
angular momentum 
$\hbox{\boldmath$J$}=\hbox{\boldmath$L$}+\hbox{\boldmath$S$}$ 
(see Fig.~1).

Next we consider the evolution of the spin vector. 
To the linear order in $S$, the equations to be solved are 
\begin{eqnarray}
 \dot \zeta^0 & = & A\zeta^1+C\zeta^2 \tilde\theta-Sv^0 \zeta^a R_a,
\nonumber \\
 \dot \zeta^1 & = & A\zeta^0+B\zeta^3+ E\zeta^2,
\nonumber \\
 \dot \zeta^2 & = & \left(C\zeta^0+D\zeta^3\right)\tilde\theta - E\zeta^1,
\nonumber \\
 \dot \zeta^3 & = & -B\zeta^1-D\zeta^2\tilde\theta-Sv^3 \zeta^a R_a. 
\end{eqnarray}
The third equation is written down explicitly as 
\begin{equation}
\dot \zeta^2=-\bar\theta\zeta_{\parallel}\kappa \sin\phi\cos\phi,
\end{equation}
with
\begin{equation}
   \kappa:=\alpha D -\beta C-\Omega_p \omega_3 r.
\end{equation}
Thus we find that
\begin{equation}
\zeta^2=-\zeta_{\perp}
+{\theta_0\zeta_{\parallel}\kappa \over 4\Omega_p}\cos 2\phi.
\end{equation}
Since the spin vector $S^a$ is itself of $O(S)$ already, 
the effect of the second term is always unimportant as long as 
we neglect corrections of $O(S^2)$ to the orbit.

The remaining three equations
determine $\alpha$, $\beta$ and $\Omega_p$. 
Corrections to $\alpha$ and $\beta$ of $O(S)$ are less interesting  
because they remain to be small however long the time passes. 
On the other hand, the correction to $\Omega_p$ will cause a big effect 
after a sufficiently long lapse of time
 because it appears in the combination of
$\Omega_p \tau$. The small phase correction will be accumulated to
become large. 
Hence, we solve $\Omega_p$ alone to the next leading order. 
Eliminating $\zeta^0$ and $\zeta^3$ from these three equations, 
we obtain
\begin{equation}
\left[(B^2-A^2)-\Omega_p^2\right]
   ={S_{\perp}\over x}
    \left({AC-BD\over r}-\Omega_p^2\omega_3 \right). 
\end{equation}
Then after a straightforward calculation, we find
\begin{equation}
\Omega_p^2={M\over r^3}\left\{
          1-{3 S_{\perp}\over r^{3/2}}
         {\pm\sqrt{M}\left(2r^2-3Mr+a^2\right)
         +ar^{1/2}(M-r)\over r^2-3Mr\pm2a\sqrt{Mr}}\right\}.
\end{equation}
As noted above, $\Omega_p$ is different from $\Omega$ for
 $S_{\perp} \not=0$.
The difference $\Omega_p-\Omega$ gives the angular velocity of the 
precession of the spin vector (see Fig.~1). 

\section{Gravitaional waves and energy loss rate}

We now proceed to the calculation of the source term in the 
Teukolsky equation and evaluate the gravitational wave flux.
For this purpose, we must write down the expression of the 
energy momentum tensor of the spinning particle explicitly. 
We rewrite the tetrad components of the energy momentum tensor 
in the following way.
\begin{eqnarray}
 T^{ab} & = &\int d\tau \left\{
   p^{(a} v^{b)} {\delta^{(4)} (x-z(\tau))\over \sqrt{-g}}
   -e^{(a}_{~\,\nu} e^{b)}_{~\rho} \nabla_{\mu} S^{\mu\nu} v^{\rho} 
   {\delta^{(4)} (x-z(\tau))\over \sqrt{-g}}\right\}
\nonumber \\
 & = & \int d\tau \left\{\left[p^{(a} v^{b)}
         +\omega_{dc}^{~\,(a} v^{b)} S^{dc} -\omega_{dc}^{~\,(a}
    S^{b)d} v^c \right] {\delta^{(4)} (x-z(\tau))\over \sqrt{-g}}
    - {1\over \sqrt{-g}}\partial_\mu\left(
    S^{\mu (a} v^{b)} \delta^{(4)} (x-z(\tau))\right)\right\}
\nonumber \\
 & =: & \mu\int d\tau \left\{A^{ab} {\delta^{(4)} (x-z(\tau))\over \sqrt{-g}} 
    + {1\over \sqrt{-g}}\partial_\mu\left(
      B^{\mu ab}\delta^{(4)}(x-z(\tau))\right)\right\}.
\end{eqnarray}
The last line is the definition of $A^{ab}$ and $B^{\mu ab}$. 

The source term of the Teukolsky equation
is expressed in terms of the components of the energy momentum 
tensor projected with respect to the complex null tetrad defined as 
\begin{eqnarray}
 \ell^{\mu} & = & \sqrt{{\Sigma\over\Delta}} \left(
             e^{~\mu}_0 +e^{~\mu}_1\right), 
 \nonumber \\
 n^{\mu} & = &  {1\over 2}\sqrt{{\Delta\over \Sigma}} \left(
             e^{~\mu}_0 -e^{~\mu}_1\right), 
 \nonumber \\
 m^{\mu} & = &  (r+ia\cos\theta)^{-1}\sqrt{\Sigma \over 2} \left(
             e^{~\mu}_2 +i e^{~\mu}_3\right).  
 \nonumber \\
\end{eqnarray}
We adopt the notation such as $T_{nn}:=n^{\mu}n^{\nu}T_{\mu\nu}$ to
denote the tetrad components.
Then the source term is given by \cite{SSTT} 
\begin{equation}
T_{\ell m\omega}=4\int d\Omega\, dt
\rho^{-5}{\bar\rho}^{-1}(B_2'+B_2'^*)
e^{-im\varphi+i\omega t}{_{-2}S^{a\omega}_{\ell m} \over \sqrt{2\pi}},
\label{teuq}
\end{equation}
where
\begin{eqnarray}
B_2' &=&-{1 \over 2}\rho^8{\bar \rho}L_{-1}[\rho^{-4}L_0
(\rho^{-2}{\bar \rho}^{-1}T_{nn})]
\nonumber \\
&- &{1 \over 2\sqrt{2}}\rho^8{\bar \rho}\Delta^2 L_{-1}[\rho^{-4}
{\bar \rho}^2 J_+(\rho^{-2}{\bar \rho}^{-2}\Delta^{-1}
T_{{\bar m}n})],
\nonumber \\
B_2'^* & =&-{1 \over 4}\rho^8{\bar \rho}\Delta^2 J_+[\rho^{-4}J_+
(\rho^{-2}{\bar \rho}T_{{\bar m}{\bar m}})]
\nonumber \\
&- & {1 \over 2\sqrt{2}}\rho^8{\bar \rho}\Delta^2 J_+[\rho^{-4}
{\bar \rho}^2 \Delta^{-1} L_{-1}(\rho^{-2}{\bar \rho}^{-2}
T_{{\bar m}n})],
\label{teu}
\end{eqnarray}
with
\begin{eqnarray}
\rho&&=(r-ia\cos\theta)^{-1},
\nonumber \\
L_j&&=\partial_{\theta}+{m \over \sin\theta}
-a\omega\sin\theta+j\cot\theta,
\nonumber \\
J_+&&=\partial_r+{iK \over \Delta},
\end{eqnarray}
and ${\bar Q}$ denotes the complex conjugate of $Q$.

As we will see shortly, the terms proportional to 
$S_{\parallel}$ in the energy momentum tensor does not contribute 
to the energy and angular momentum fluxes at the linear order 
in $S$. In other words, the energy and angular momentum fluxes are the
same for all the orbits having the same $S_\perp$, irrespective of
the value of $S_\parallel$.
Thus, we ignore these terms in the following discussion. 
Further we recall that the particle can stay in the equatorial plane 
if $S_{\parallel}=0$.  Hence we fix $\theta=\pi/2$ in the following 
calculations. 

Using the formula (\ref{Rinfty}), 
we obtain the amplitude of gravitaional waves at infinity as 
\begin{equation}
\tilde Z_{\ell m\omega} =
  \tilde Z^{nn}_{\ell m\omega}
 +\tilde Z^{\bar m n}_{\ell m\omega}
 +\tilde Z^{\bar m \bar m}_{\ell m\omega},
\end{equation}
where
\begin{eqnarray}
 \tilde Z^{nn}_{\ell m\omega} & = & 
 {i\sqrt{2\pi}\over \omega B^{in}_{\ell m\omega}}
   \delta\left(\omega-m\Omega\right)
   \left({dt\over d\tau}\right)^{-1} \left[
   A_{nn}-i\omega B^t_{nn}
   +imB^{\varphi}_{nn}-B^r_{nn}{\partial\over \partial r}\right]
\nonumber \\
  && 
   \times\left[L^{\dag}_1 \rho^{-4}\left(L^{\dag}_2 \rho^3 
    {}_{-2}S^{a\omega}_{\ell m}\right)\right]_{\theta=\pi/2}
    {1\over r\Delta} R^{in}_{\ell m\omega}
    \Biggr\vert_{r=r_0},
\nonumber \\     
 \tilde Z^{\bar m n}_{\ell m\omega} & = & 
 {i\sqrt{\pi}\over \omega B^{in}_{\ell m\omega}}
   \delta \left(\omega-m\Omega\right)
   \left({dt\over d\tau}\right)^{-1} \left[
   A_{\bar m n}-i\omega B^t_{\bar m n}
   +imB^{\varphi}_{\bar m n}-B^r_{\bar m n}
   {\partial\over \partial r}\right]
\nonumber \\
  && 
   \times\left(L^{\dag}_2 {}_{-2}S^{a\omega}_{\ell m}
   \right)_{\theta=\pi/2}{1\over \sqrt{\Delta}}
   \left[2{\partial\over \partial r}-{2iK\over\Delta}-{4\over r}\right]
     R^{in}_{\ell m\omega}\Biggr\vert_{r=r_0},
\nonumber \\     
 \tilde Z^{\bar m \bar m}_{\ell m\omega} & = & 
 {i\sqrt{\pi}\over \omega B^{in}_{\ell m\omega}}
   \delta\left(\omega-m\Omega\right)
   \left({dt\over d\tau}\right)^{-1} \left[
   A_{\bar m \bar m}-i\omega B^t_{\bar m \bar m}
   +imB^{\varphi}_{\bar m \bar m}-B^r_{\bar m \bar m}
   {\partial\over \partial r}\right]
\nonumber \\
  && 
   \times\left({}_{-2}S^{a\omega}_{\ell m}\right)_{\theta=\pi/2}
   \left[{\partial^2\over \partial r^2}
   -2\left({1\over r}+{iK\over\Delta}\right){\partial\over \partial r}
    -\left({iK\over\Delta}\right)_{,r}+{2iK\over\Delta r}
    -{K^2\over\Delta^2}\right]R^{in}_{\ell m\omega}\Biggr\vert_{r=r_0},
\label{Ztilde}
\end{eqnarray}
and
\begin{eqnarray}
 A_{nn} & = & {1 \over 4}{1\over 1-x^2}\left\{
     1 - S_{\perp} \left( (2\omega_1 +\omega_3)x +\omega_2
     \right)\right\},
\nonumber \\
 B^{\mu}_{nn} & = & {1\over 4r}S_{\perp}{1\over 1-x^2}
   \left( {r^2 + a^2\over\sqrt{\Delta}}x+a,
       ~-\sqrt{\Delta}x,~ 0, ~{a\over\sqrt{\Delta}}x+ 1\right),
\nonumber \\
 A_{\bar m n} & = & {i \over 4\sqrt{2}}
    {1\over 1-x^2}\left\{
     2x -S_{\perp} \left(\omega_1 x^2 -4\omega_2 x -\omega_3
     \right)\right\},
\nonumber \\
 B^{\mu}_{\bar m n} & = & {i \over 4\sqrt{2}r}S_{\perp}
   {1\over 1-x^2} \left( {r^2 +a^2\over\sqrt{\Delta}}x^2+ax,
       ~-\sqrt{\Delta}(1+x^2), ~0, ~{a\over\sqrt{\Delta}}x^2+ x\right),
\nonumber \\
 A_{\bar m \bar m} & = & -{1\over 2}
    {1\over 1-x^2}\left\{
     x^2 +S_{\perp} \left(\omega_2 (1+2x^2) +\omega_3 x
     \right)\right\},
\nonumber \\
 B^{\mu}_{\bar m \bar m} & = & {1\over 2r}S_{\perp} {1\over 1-x^2}
   \left(0, ~\sqrt{\Delta} x, ~0, ~0\right),
\end{eqnarray}

and 
\begin{equation}
L_j^{\dag}=\partial_{\theta}-{m \over \sin\theta}
+a\omega\sin\theta+j\cot\theta.
\end{equation}
The Lorentz factor $dt/ d\tau$ which appears in Eqs.~(\ref{Ztilde}) 
can be calculated from Eqs.~(\ref{v0v3}) as
\begin{equation}
 {dt\over d\tau} = {1\ \over r\sqrt{1-x^2}}\left(
	   ax+ {r^2+a^2\over\sqrt\Delta}\right).
\end{equation}

When the orbit is quasi-periodic,
the Fourier component of gravitational waves do not have a
continuous spectrum but takes the form,
\begin{equation}
\tilde Z_{\ell m\omega}=\sum_{n}\delta(\omega-\omega_n)Z_{\ell m\omega_n}.
\end{equation}
Then the time averaged energy flux is given by the formula\cite{SSTT},
\begin{equation}
\left\langle {dE\over dt}\right\rangle_{GW} =
 \sum_{\ell,m,n}{\left\vert Z_{\ell m\omega_n}\right\vert^2
             \over 4\pi \omega_n^2}
 =:\sum_{\ell,m,n}\left({dE\over dt}\right)_{\ell m n}.  
\end{equation}
The the $z$-component of the angular momentum flux is
also given by the similar formula,
\begin{equation}
\left\langle {dJ_z\over dt}\right\rangle_{GW} =
 \sum_{\ell,m,n}{m \left\vert Z_{\ell m\omega_n}\right\vert^2
             \over 4\pi \omega_n^3}
 =:\sum_{\ell,m,n}\left({dJ_z\over dt}\right)_{\ell m n}.
\end{equation}
In the present case of circular orbits in the equatorial plane,
the index $n$ degenerates to the angular index $m$ and
$\omega_n$ is simply given by $m\Omega$ ($n=m$).
Hence we eliminate the index $n$ in the following discussion.

Here we mention the effect of non-zero $S_\parallel$.
If we recall that all the terms which are proportional to 
$S_{\parallel}$ have the time dependence of $e^{\pm i\Omega_p \tau}$, 
we see that they give the contribution to the side bands.
That is to say, their contributions in $\tilde Z_{\ell m\omega}$ are 
all proportional to $\delta(\omega-m\Omega\pm\Omega_p)$. 
Then, since the energy and angular momentum fluxes are quadratic in
$Z_{\ell m\omega_n}$, they are not affected by the presence of
$S_\parallel$ as long as we are working only up to the linear order in 
$S$.

In order to express the post-Newtonian corrections to the 
energy flux, we define $\eta_{\ell m\omega}$ as
\begin{equation}
 \left({dE\over dt}\right)_{\ell m}
   =:{1\over 2}\left({dE\over dt}\right)_{N}\eta_{\ell m},
\end{equation}
where $(dE/ dt)_{N}$ is the Newtonian quadrupole formula
\begin{equation}
 \left({dE\over dt}\right)_{N}
  ={32\mu^2 M^3\over 5 r^5}
   =:{32\over 5}\left({\mu\over M}\right)^2 v^{10}.
\end{equation}
We calculate $\eta_{\ell m}$ up to 2.5PN order, i.e., to $O(v^5)$. 
The result is
\begin{eqnarray}
 \eta_{2\pm2} & = & 1-{107\over 21} v^2 
         +\left(4\pi -6q -{19 \over 3}\hat s \right) v^3
	     +\left({4784\over 1323} +2 q^2 +9q\hat s\right) v^4
	     +\left(-{428\over 21}\pi +{4216\over 189} q
		   +{2134\over 63} \hat s\right) v^5,
\nonumber \\
 \eta_{2\pm 1} & = & {1\over 36} v^2 
         +\left(-{1\over 12}q +{1 \over 12}\hat s \right) v^3
         +\left(-{17\over 504} +{1\over 16} q^2 
                          -{1\over 8}q\hat s\right) v^4
          +\left({1\over 18}\pi -{793\over 9072} q 
	      -{535\over 1008} \hat s\right) v^5,
\nonumber \\
 \eta_{3\pm 3}& = & {1215\over 896} v^2 
	     -{1215\over 112} v^4
	     +\left({3645\over 448}\pi -{1215\over 112} q
	       -{10935\over 896} \hat s\right) v^5,
\nonumber \\
 \eta_{3\pm 2}& = & 
         {5\over 63} v^4
         +\left(-{40\over 189} q
         +{20\over 63} \hat s\right) v^5,
\nonumber \\
 \eta_{3\pm 1}& = & {1\over 8064} v^2 
         -{1\over 1512} v^4
         +\left({1\over 4032}\pi -{17\over 9072} q
           -{1\over 8064} \hat s\right) v^5,
\nonumber \\
 \eta_{4\pm 4}& = & 
	     {1280\over 567} v^4,
\nonumber \\
 \eta_{4\pm 2}& = & 
	     {5\over 3969} v^4,
\end{eqnarray}
where $q:=a/M$ and $\hat s:=S_{\perp}/M$.
The rest of $\eta_{\ell m}$ are all of higher order. 
We should mention that if we regard the spinning particle as 
a model of a black hole or neutron star, 
$S$ is of order $\mu$. Therefore the corrections due to 
$S$ are generally small compared with the $S$-independent terms in the
test particle limit $\mu/M\ll1$.

Putting all together, we obtain to 2.5PN order, 
\begin{eqnarray}
 \left\langle dE\over dt\right\rangle_{GW}= \left(dE\over dt\right)_N
     &&\Biggl[ 1-{1247\over 336} v^2 
	     +\left(4\pi -{73\over 12}q -{25 \over 4}\hat s \right) v^3
\nonumber \\
	     && +\left(-{44711\over 9072} +{33\over 16} q^2 
		  +{71\over 8} q\hat s\right) v^4
	     +\left(-{8191\over 672}\pi +{3749\over 336} q
		   +{2403\over 112}\hat s\right) v^5\Biggr].
\end{eqnarray}
Since $v$ is defined in terms of the coordinate radius of the orbit,
the expansion with respect to $v$ does not have a clear
 gauge-invariant meaning. In particular, for the purpose of the
comparison with the standard post-Newtonian calculations,
 it is better to write the result 
by means of the angular velocity observed at the infinity. 
Using the post-Newtonian expansion of Eq.~(\ref{Omegaori}),
\begin{equation}
 M\Omega = v^{3}
   \left(1-\left({3\over 2}\hat s+q \right)v^{3}
    +{3\over 2}q\hat s v^{4}
    +O\left(v^6\right)\right), 
\end{equation}
the above result can be rewritten as 
\begin{eqnarray}
 \left\langle dE\over dt\right\rangle_{GW}
= \left(\widetilde{dE\over dt}\right)_N
     &&\Biggl[ 1-{1247\over 336} (M\Omega)^{2/3} 
     +\left(4\pi -{11\over 4}q -{5 \over 4}\hat s \right) (M\Omega)
\nonumber \\
     && +\left(-{44711\over 9072} +{33\over 16} q^2 
	  +{31\over 8} q\hat s\right) (M\Omega)^{4/3}
     +\left(-{8191\over 672}\pi +{59\over 16} q
	   -{13\over 16} \hat s\right) (M\Omega)^{5/3}\Biggr],
\label{dE/dt}
\end{eqnarray}
where 
\begin{equation}
 \left(\widetilde{dE\over dt}\right)_N :=
 {32\over 5}\left({\mu\over M}\right)^2 (M\Omega)^{10/3}.
\end{equation}
Since there is no side-band contribution in the present case,
the angular momentum flux is simply given by 
$\left\langle dJ_z/dt\right\rangle_{GW}=\Omega^{-1}
\left\langle dE/dt\right\rangle_{GW}$.
The result (\ref{dE/dt}) is consistent with the one obtained by the 
standard post-Newtonian approach\cite{kw2,kidd} to the 2PN order 
in the limit $\mu/M \rightarrow 0$.
The $\hat s$-dependent term of order $(M\Omega)^{5/3}$ is the 
one which is newly obtained here. 

\section{radiation reaction}
In this section, we consider the effect of radiation reaction
on the orbit by equating the gravitational
energy and angular momentum fluxes with
their loss rates of the system.

\subsection{Conserved Quantities}

Here we consider the conserved quantities which are the 
first integrals of the equations of motion. 
First we give two conserved quantities which follow from the 
Killing vectors of the Kerr spacetime. 
The timelike Killing vector is given by 
\begin{equation}
 \xi_{\mu}=\sqrt{{\Delta\over\Sigma}} ~e^0_{~\mu}
    +{a\sin\theta \over \sqrt{\Sigma}} ~e^3_{~\mu}, 
\end{equation}
and its derivative is 
\begin{equation}
 \xi_{\mu;\nu} = -{2M(r^2 -a^2\cos^2\theta)\over \Sigma^2} 
  ~e^1_{~[\mu} e^0_{~\nu]} 
  +{4Mar\cos\theta\over\Sigma^2} ~e^3_{~[\mu} e^2_{~\nu]}.
\end{equation}
The rotational Killing vector is given by 
\begin{equation}
 \chi_{\mu}=a\sin^2\theta\sqrt{{\Delta\over\Sigma}} ~e^0_{~\mu}
    +{(r^2+a^2)\sin\theta \over \sqrt{\Sigma}} ~e^3_{~\mu}, 
\end{equation}
and its derivative is 
\begin{eqnarray}
 \chi_{\mu;\nu} = && -{2a\sin^2\theta\over \Sigma^2}
  \left((r-M)\Sigma +2Mr^2\right) e^1_{~[\mu} e^0_{~\nu]} 
\nonumber \\
 && -{2a\sqrt{\Delta}\sin\theta\cos\theta\over \Sigma}
 ~e^2_{~[\mu} e^0_{~\nu]} 
 -{2r\sin\theta\sqrt{\Delta}\over \Sigma}
  ~e^1_{~[\mu} e^3_{~\nu]}
\nonumber \\
 && +{2\cos\theta\over \Sigma^2}\left(
 a^2\sin^2\theta\Delta -\left(r^2+a^2\right)^2\right)
 e^2_{~[\mu} e^3_{~\nu]}.
\end{eqnarray}
Then following the discussion around Eq.~(\ref{cons}), 
we can construct the conserved quantities 
describing the energy and the $z$-component 
of the angular momentum from these Killing vectors. 
They are given by
\begin{eqnarray}
{E\over\mu} & 
  := & u^{\mu} \xi_{\mu} -{1\over 2\mu}S^{\mu\nu} \xi_{\mu;\nu}
\nonumber \\
& = & \sqrt{{\Delta\over\Sigma}}u^0 
+{a\sin\theta\over\sqrt{\Sigma}}u^3
+{M(r^2-a^2\cos^2\theta)\over\Sigma^2}{S^{10}\over\mu}
+{2Mar\cos\theta\over\Sigma^2}{S^{23}\over\mu}\,,
\nonumber \\
{J_z\over \mu} & 
 := & u^{\mu} \chi_{\mu} -{1\over 2\mu}S^{\mu\nu} \chi_{\mu;\nu}
\nonumber \\
& = & a\sin^2\theta \sqrt{{\Delta\over\Sigma}}u^0 
+{(r^2+a^2)\sin\theta\over\sqrt{\Sigma}}u^3
+{a\sin^2 \theta\over\Sigma^2}\left((r-M)\Sigma +2Mr^2\right)
{S^{10}\over\mu}
\nonumber \\
&&+{a\sqrt{\Delta}\sin\theta\cos\theta\over\Sigma}{S^{20}\over\mu}
+{r\sqrt{\Delta}\sin\theta\over\Sigma}{S^{13}\over\mu}
-{\cos\theta\over\Sigma^2}\left(a^2\sin^2\theta\Delta 
-\left(r^2+a^2\right)^2\right) {S^{23}\over\mu}\,,
\label{consel}
\end{eqnarray}
Since terms such as $\cos\theta S^{23}$ and  
$\cos\theta S^{20}$ are higher order in $\tilde \theta$, 
we neglect them in the following discussion. 

For a spinless particle, there exists one more conserved quantity on the Kerr
spacetime, known as the Carter constant\cite{carter}. 
It is associated with the 
Killing tensor $K_{\mu\nu}$ which satisfies $K_{(\mu\nu;\sigma)}=0$.
However, for a spinning particle, no such conserved quantity has been known.
Nevertheless, one can show that there exists an approximately conserved 
quantity which corresponds to the Carter constant for a spinless particle.
It is constructed as follows.

In addition to the Killing vectors, the Kerr spacetime has an antisymmetric 
Killing Yano tensor,
\begin{equation}
 f_{\mu\nu}= 2a\cos\theta ~e^1_{~[\mu} e^0_{~\nu]} 
      +2r ~e^2_{~[\mu} e^3_{~\nu]}, 
\end{equation}
which satisfies 
\begin{equation}
 f_{\mu(\nu;\sigma)}=0.
\end{equation}
Note that from $f_{\mu\nu}$, 
we can construct the Killing tensor $K_{\mu\nu}$ as 
\begin{equation}
K_{\mu\nu}=f_{\mu\sigma}f_{\nu}^{~\sigma}=r^2g_{\mu\nu}
   +2\Sigma l_{(\mu}n_{\sigma)}. 
\end{equation}
When there is a Killing Yano tensor, the system possesses 
a quantity whose $\tau$-derivative is of $O(S^2)$, hence conserved
to the linear order in $S$.
Introducing the totally antisymmetric tensor,
\begin{eqnarray}
 f_{\mu\nu\sigma} & := & f_{\mu\nu;\sigma}
 \nonumber \\
 &=& 6\left({a\sin\theta\over\sqrt{\Sigma}}
     ~e^0_{~[\mu} ~e^1_{~\nu} e^2_{~\sigma]}
     +\sqrt{{\Delta\over\Sigma}}
       ~e^1_{~[\mu} ~e^2_{~\nu} e^3_{~\sigma]}\right),
\end{eqnarray}
the approximate conserved quantity is expressed by \cite{SUSY} 
\begin{eqnarray}
{Q\over \mu^2} & 
:= & {1\over 2}f_{\mu\sigma} f_{\nu}^{~\sigma} u^{\mu} u^{\nu}
   -u^{\mu} {S^{\rho\sigma}\over\mu}
\left(f^{\nu}_{~\sigma} f_{\mu\rho\nu} 
      -f_{\mu}^{~\nu}f_{\rho\sigma\nu}\right)
 \nonumber \\
  & = & {1\over 2} 
  \left(\Sigma\left((u^0)^2-(u^1)^2\right)-r^2\right)
 \nonumber \\
  &&  -{a\sin\theta\over\mu\sqrt{\Sigma}}\left\{
     r\left(u^0 S^{13}+u^1 S^{30}-2u^3 S^{10}\right)      
     +a\cos\theta u^3 S^{23}\right\}
 \nonumber \\
  &&  -{1\over\mu}\sqrt{{\Delta\over\Sigma}}\left\{
     a\cos\theta\left(u^2 S^{30} -u^3 S^{20}+2u^0 S^{23}\right)      
     -r u^0 S^{10}\right\}. 
\end{eqnarray}
The quantity corresponding to the Carter constant is 
\begin{equation}
 C:=2Q-(J_z -a E)^2. 
\end{equation}
For the case of circular orbits under consideration, we find
\begin{equation}
{C\over\mu^2}=-2aS_\perp\,.
\label{carterconst}
\end{equation}

\subsection{Frequency Shift due to Radiation Reaction}

The orbit of a non-spinning particle is completely determined by the three 
constants of motion, $E$, $J_z$ and $C$.
As mentioned above the counterpart of $C$ for a spinning 
particle also exists in the linear order of $S$.
This means that we need to calculate the radiation 
reaction to these quantities to obtain the orbital evolution of a spinning 
particle. In both non-spinning and spinning 
cases, the radiation reaction to $E$ and $J_z$ can be 
evaluated by equating their loss rates with the corresponding fluxes 
emitted by gravitational waves.
However, we do not know how to determine the reaction to $C$
from the asymptotic behavior of emitted gravitational waves at infinity.
Furthermore, there are also spin degrees of freedom in the present case. 
It is not clear at all how to evaluate the backreaction 
to the spin. In order to fully understand the radiation 
reaction, it will be necessary to derive some regularized radiation reaction 
force which acts on the particle.  
However, it is beyond the scope of the present paper. 

Although the rigorous evaluation of the change rate of spin 
seems formidable, there exists 
an order-of-magnitude estimate by Apostolatos et al.\cite{apo}, 
in which they calculate a torque acted inside a spinning star due to the 
radiation reaction force. According to their 
estimate, the rate of change of the spin is $O(v^{10})$ smaller than 
that of the orbital angular momentum.  
This means we can safely ignore the time variation of the spin if $v \ll 1$. 
Since we are interested in the case $v \ll 1$ in this paper,
we may then assume that the circular orbit obtained in section 4
remains circular with smaller $r$ but 
with the same values of $S_{\perp}$ and $S_{\parallel}$ under
the radiation reaction. Note that this implies the approximate
Carter constant, Eq.~(\ref{carterconst}), is conserved under the
radiation reaction.

If this assumption is correct, we can obtain an evolution sequence
of the orbits under the radiation reaction. Namely the orbit is
quasi-circular and slowly spiralling in with constant $S_\perp$
and $S_\parallel$.
Although not sufficient,
we can consider the necessary condition for this assumption to be true by
examining the consistency with the energy and angular  
momentum loss rates evaluated in terms of their fluxes emitted by gravitational
waves. Here the consistency means that the relation
\begin{equation}
 \left\langle{dE\over dt}\right\rangle_{GW}=
{\delta E\over \delta J_z}\left\langle{dJ_{z}\over dt}\right\rangle_{GW}\,,
\label{balance}
\end{equation}
holds, where the variations $\delta E$ and $\delta J_z$ 
are taken keeping the orbit circular 
with fixed $S_{\perp}$ and $S_{\parallel}$. 
As noted before, since there are no side-band contributions to the energy and 
angular momentum fluxes in the present case, 
we have $\langle dE/dt\rangle_{GW}=\Omega\langle dJ_z/dt\rangle_{GW}$. 
Therefore the condition (\ref{balance}) is equivalent to 
\begin{equation}
{\delta E\over \delta J_z}=\Omega\,.
\label{tbp}
\end{equation}
In the following, we prove that Eq.~(\ref{tbp}) indeed holds. 
Since the $S_{\parallel}$ does not appear in the expressions for $E$, $J_z$, 
$\langle dE/dt\rangle_{GW}$ and $\langle dJ_z/dt\rangle_{GW}$ 
to the linear order in $S$, 
we only have to examine the situation with $S_{\parallel}=0$.

We introduce the function,
\begin{equation}
 V(E,J_z,r,S_{\perp}):=-g^{tt}u_t u_t -2g^{t\varphi}u_t u_{\varphi}
           -g^{\varphi\varphi}u_{\varphi} u_{\varphi}-1, 
\label{Vdef}
\end{equation}
which is guaranteed to be non-negative and it becomes zero 
when the orbit has no radial or vertical motion, i.e., 
$v^1=v^2=0$.
If we set $E$ and $J_z$ to the values for a circular orbit with 
$r=r_0$, we have $V=0$ at $r=r_0$ and $V>0$ at any other points
near $r=r_0$.
Hence ${\partial V/\partial r}\vert_{r=r_0}=0$.

The momentum components $-\mu u_t$ and $\mu u_{\varphi}$ are 
different from the conserved 
energy and angular momenta due to the presence of spin. 
{}From Eqs.~(\ref{consel}), we see
\begin{eqnarray}
-\mu u_t & = & E-
\mu S_{\perp} \varepsilon\left(u^0(E,J_z),u^3(E,J_z),r\right),
\nonumber \\
\mu u_{\varphi} & = & J_z-
\mu S_{\perp} j\left(u^0(E,l_z),u^3(E,l_z),r\right),
\end{eqnarray}
where
\begin{eqnarray}
 \varepsilon(u^0,u^3,r) & = & {M\over r^2} u^3,
\nonumber \\
 j(u^0,u^3,r) & = & -{a\over r^2}(M+r)u^3 
                  -{\sqrt{\Delta}\over r}u^0.
\end{eqnarray}
Inserting these expressions to Eq.~(\ref{Vdef}),
taking the variation of $V$ while keeping the orbit circular 
with fixed spin, and recalling ${\partial V/\partial r}\vert_{r=r_0}=0$,
we obtain the relation between $\delta E$ and 
$\delta J_z$ as
\begin{eqnarray}
 \mu \delta V & = &2 u^t \left[\delta E -\mu S_{\perp}
           \left({\partial \varepsilon\over \partial E} 
                 \delta E
        + {\partial \varepsilon\over \partial J_z} \delta J_z\right)
                \right]
          -2 u^{\varphi} \left[\delta J_z-\mu S_{\perp}
          \left({\partial j\over \partial E} \delta E
                + {\partial j\over \partial J_z} \delta J_z\right)\right]
 \nonumber \\        
 & =& 0. 
\end{eqnarray}
This gives 
\begin{equation}
 {\delta E\over \delta J_z}={u^{\varphi}\over u^t}
      \left\{
             1+\mu S_{\perp} \left[{1\over u^{\varphi}}
             \left(-{\partial \varepsilon\over \partial J_z} u^t
             +{\partial j\over \partial J_z} u^{\varphi}\right)
             +{1\over u^t}
             \left(-{\partial \varepsilon\over 
                                \partial E} u^t
             +{\partial j\over \partial E} 
                  u^{\varphi}\right)\right]
                  +O(S^2)\right\}.
\label{cconv}
\end{equation}
By using the following relations,
\begin{eqnarray}
 u^0 & = &{1\over r\sqrt{\Delta}}\left((r^2+a^2)E-a J_z\right),
\nonumber \\
 u^3 & = &{1\over r}\left(J_z -a E\right),
\nonumber \\
 u^t & = & {1\over \sqrt{\Delta}}{r^2+a^2\over r} u^0 +{a\over r} u^3,
\nonumber \\
 u^{\varphi} & = & {1\over \sqrt{\Delta}}{a\over r}u^0+{1\over r}  u^3,
\end{eqnarray}
which hold in the lowest order in $S$,
it is easy to verify that the terms in the square parentheses in the 
right-hand side of Eq.~(\ref{cconv}) become of $O(S^2)$ or higher
at $r=r_0$. 
Thus Eq.~(\ref{tbp}) is shown to hold to the linear order in $S$ and
our assumption of stability of the quasi-circular orbit is found to be
consistent.

Under the assumption that the orbit remains quasi-circular 
with fixed spin, we can evaluate the frequency shift due 
to the radiation reaction by
\begin{equation}
  {d\Omega\over dt}=-\left({dE\over d\Omega}\right)^{-1}
                   \left\langle{dE\over dt}\right\rangle_{GW}. 
\end{equation}
The post-Newtonian expansion of this quantity is calculated to become 
\begin{eqnarray}
 {d\Omega\over \Omega^2 dt} &=&
   {5\over 96}{\mu\over M}
     (M\Omega)^{5/3}\Biggl[1-{743\over 336}(M\Omega)^{2/3}
               +\left(4\pi-{113\over 12}q-{25\over 4}\hat s\right)(M\Omega)
\nonumber \\
  &&+\left({34103\over 18144}+{81\over 16}q^2+{79\over 8}q\hat s\right)
                                               (M\Omega)^{4/3}
    +\left(-{4159\over 672}\pi-{31319\over 1008}q-{809\over 84}\hat s\right)
                                               (M\Omega)^{5/3}\Biggr]. 
\end{eqnarray}

\section{summary and discussion}

In this paper we have investigated the gravitational waves 
emitted by a spinning particle in circular orbits around a rotating 
black hole. First we have solved the equations of motion of 
a spinning particle in the Kerr spacetime,  
assuming the spin of the particle is small and 
the orbit is close to the equatorial plain.  
Applying the Teukolsky formalism of the black hole perturbation,  
we have then calculated the first order corrections due to spin 
to the energy flux up to 2.5PN order. 
The effect of spin is always small $(\sim O(\mu/ M))$ 
compared with the spin-independent contributions
if we take the limit $\mu/M\rightarrow 0$. 
However, the result will be a useful guideline to the 
standard post-Newtonian calculations because 
our approach is totally different from the standard post-Newtonian
approach and gives 
the leading spin-dependent terms in the $\mu/M$ 
expansion.
Up to 2PN order, 
our results are in complete agreement with the 
previous ones obtained by the standard post-Newtonian method.
The spin-dependent term at 2.5PN order, which we have newly obtained,
will be verified by the standard post-Newtonian approach
in future.

In this paper we have restricted our analysis to 
a class of orbits which are circular and stays in 
the equatorial plane if the spin vector is orthogonal to it.
In this case, the assumption that the orbit remains in this class 
under radiation reaction has been found to be consistent 
with the energy and the angular momentum loss rates evaluated from 
the gravitational wave flux at infinity.
Although we have not considered it here, it seems
possible to incorporate the orbital inclination.
However, if the present restriction is relaxed, the orbit may become 
too complicated and the same technique used here may not work well. 
As long as one considers a spinless particle, 
the orbit is parameterized by the three conserved quantities, 
i.e., the energy, the $z$-component of the angular momentum
and the Carter constant. In that case, the particle will sweep
through a certain region of the phase space restricted by
these conserved quantities 
sufficiently fast compared with the time scale of the radiation reaction.  
Hence the quasi-periodicity will be a good approximation.
On the other hand,  for a spinning particle,
there are not sufficient number of conserved quantities
to confine the orbit to a restricted region of the phase space. 
This means that the orbit is not guaranteed to be quasi-periodic. 
When we try to make a better template to be used for 
interferometric gravitational wave detectors in future
by taking account of the effect of spin,
this point may cause a difficult problem. 

\acknowledgments
We thank H. Asada, T. Nakamura and H. Tagoshi for helpful disucussions. 
We also thank M. Tanimoto for informing us of the reference [25].
This work was supported in part by the Monbusho Grant-in-Aid for
Scientific Research Nos. 5326, 5423, and 07740355. 


\appendix
\section{}
In this appendix, we present an exact solution of 
the equations of motion in the case when the spin vector $S^a$ is 
parallel or anti-parallel to the rotation axis of the Kerr black hole. 
We assume the orbit is circular and lies on the equatrial plane. 
Thus the only non-vanishing component
of the spin vector is $S \zeta^2=-S_{\perp}$ and we set 
$v^1=v^2=0$ exactly.
Further we assume $u^1=u^2=0$, which will be found to be consistent.

With these assumptions,
the only non-vanishing component of $R^a$ is 
\begin{equation}
SR^1=-S_{\perp}{M\over r^3}\left(
     2v^0 u^3+v^3u^0)\right).
\end{equation}
Then all but one of the equations of motion 
are trivially satisfied and
the remaining non-trivial equation is 
\begin{equation}
 \dot u^1 = \omega_1 v^0 u^0 + \omega_2 
           (v^0 u^3 + v^3 u^0)+ \omega_3 v^3 u^3 
          +S_{\perp}{M\over r^3} (2v^0 u^3 +v^3 u^0)=0. 
 \label{p1dot}
\end{equation}
Besides the equations of motion there are constraint equations 
to be satisfied. We list them below. 
\begin{enumerate}
\item
>From the time derivative of the center of mass condition $S^{ab}u_b=0$, 
we have the relation between $u^a$ and $v^a$ as
\begin{equation}
 u^a=v^a -{S^{ab}\over\mu}SR_{b}\,,
 \label{pvrel}
\end{equation}
which is equivalent to Eq.~(\ref{vprelation}).
This gives
\begin{eqnarray}
 u^0 & = &  v^0 - {MS_\perp^2\over r^3}(2 v^0u^3+v^3u^0)u^3, 
 \label{pvrel0}\\
 u^3 & = &  v^3 - {MS_\perp^2\over r^3}(2 v^0u^3+v^3u^0)u^0. 
 \label{pvrel3}
\end{eqnarray}
\item 
The mass conservation $u_a u^a=-1$ gives
\begin{equation}
 \left( u^0 \right)^2 -\left( u^3 \right)^2 = 1.
 \label{mcons} 
\end{equation}
\item 
The normalization condition $u_a v^a=-1$ of $v^a$ is
\begin{equation}
 u^0 v^0 -u^3 v^3 =1.
 \label{vnorm}
\end{equation}
\end{enumerate}
We may consider $S_\perp$ and $r$ as freely specifiable variables.
Then the variables to be determined are $v^0$, $v^3$, $u^0$ and $u^3$.
However there are five equations to be satisfied, i.e.,
Eqs.~(\ref{p1dot}), (\ref{pvrel0}) $\sim$ (\ref{vnorm}). 
Hence one might think the system is overdetermined and inconsistent.
Fortunately this is not the case
because Eq.~(\ref{vnorm}) is guaranteed to hold 
by Eqs.~(\ref{pvrel0}), (\ref{pvrel3}) and (\ref{mcons}), 
as can be seen by contracting Eq.~(\ref{pvrel}) with $u_a$.
Thus our assumptions, in particular $u^1=u^2=0$, turn out to be
consistent.

In order to solve the above set of equations, 
we introduce the new variables,
\begin{equation}
 x_v={v^3\over v^0}\quad{\rm and}\quad x_u={u^3\over u^0}\,.
\end{equation}
In terms of $x_v$ and $x_u$, Eq.~(\ref{p1dot}) is rewritten as
\begin{equation}
 a^2-Mr+a\sqrt{\Delta}(x_v+x_u)+\Delta x_v x_u 
 +\sigma M\sqrt{\Delta} (2x_u + x_v)=0,
 \label{eq:f1}
\end{equation}
where
\begin{equation}
 \sigma:={S_{\perp}\over r}={M\over r}\hat s\,.
\end{equation}
Multiplying Eq.~(\ref{pvrel0}) by $u^3$ and Eq.~(\ref{pvrel3}) by $u^0$,
equating the right-hand sides of them and dividing it by $u^0v^0$, we obtain
\begin{equation}
 x_v -{M\over r}\sigma^2(2 x_u +x_v)(u^0)^2=
 x_u -{M\over r}\sigma^2(2 x_u +x_v)(u^3)^2.
 \label{eq:f23}
\end{equation}
Using Eq.~(\ref{mcons}), one readily sees this reduces to
\begin{equation}
 x_v -x_u-{M\over r}\sigma^2(2x_u+x_v)=0.
 \label{eq:xvxp}
\end{equation}
Thus we have obtained the coupled equations 
(\ref{eq:f1}) and (\ref{eq:xvxp}) for $x_v$ and $x_u$.
They are solved to give
\begin{eqnarray}
 x_v & = & {-(2r a+3Mr\sigma +aM\sigma^2) \pm 
 \sqrt{4M r^3 +12aM r^2 \sigma +13 M^2 r^2 \sigma^2 
  +6 aM^2r \sigma^3 -8M^3r\sigma^4+9a^2M^2 \sigma^4}
 \over 2\sqrt{\Delta}(r-M\sigma^2)}, \nonumber \\
 x_u & = & {r -M\sigma^2 \over r +2M\sigma^2}x_v.
\label{xuxvsol}
\end{eqnarray}
Then from Eqs.~(\ref{mcons}) and (\ref{vnorm}),
$u^0$, $u^3$, $v^0$ and $v^3$ are found as
\begin{eqnarray}
u^0&=&{1\over\sqrt{1-x_u^2}}\,,\quad u^3={x_u\over\sqrt{1-x_u^2}}\,,
\nonumber\\
v^0&=&{\sqrt{1-x_u^2}\over1-x_vx_u}\,,
\quad v^3={x_v\sqrt{1-x_u^2}\over1-x_vx_u}\,.
\end{eqnarray}
We note that the terms inside the square root of the expression for
$x_v$ in Eq.~(\ref{xuxvsol}) is not positive definite. Thus
circular orbits do not exist for very large values of $\sigma$
 ($=S_\perp/r$). However, they always exist for physically reasonable
 values of $\sigma$, i.e., for $\sigma<M/r<1$. 
We also note that the $+$ ($-$) sign in front of the square root 
corresponds to a co-rotating (counter-rotating) orbit if we restrict 
the range of $a$ to be non-negative, i.e., $0\leq a<M$. 
On the other hand, if we extend the range of $a$ to
$-M<a<M$, the $\pm$ signs become redundant. Here we take the latter
option and take the $+$ sign.

Then a matter of interest is the stability of these circular orbits.
In Fig.~2, we show the contours of radii of 
the inner-most stable circular orbits on the $(a,S_\perp)$-plane.
The contour of $r=6M$ passes through the $(a,S_\perp)=(0,0)$,
which is the well-known minimum radius for a spinless particle in the
Schwarzchild spacetime.
One readily notices that the minimum radius increases as $a$
increases for a fixed $S_\perp$ and it is larger for larger $S_\perp$.
Another interesting feature is that the minimum radius approaches
$r=M$ in the limit $a\to-M$ irrespective of the values of $S_\perp$.
Although this latter feature can be explained only in a fully
relativistic context,
the main feature of the contours can be understood as a consequence of 
the spin-orbit coupling, which is the dominant effect in a mildly 
relativistic situation. It is repulsive when the spin and orbital
angular momentum vectors are parallel and attractive when they are 
anti-parallel.
Now, if the contribution of the particle's spin to the spin-orbit 
interaction could be neglected, the contours of the minimum radii 
would be parallel to the $S_\perp$-axis, with increasing minimum radii
for larger $a$. On the other hand, if the particle were another Kerr
black hole with the same mass $M$, spins of the black hole and the
particle would contribute to the spin-orbit interaction in an exactly
symmetric way, and the contours would be straight lines at $45^\circ$
downward in the right direction. 
In reality, neither the spin of the particle can be neglected nor
its contribution is as large as that of the black hole.
This approximately explains the feature of the contours.

\par\bigskip\par
\begin{center}
\bf FIGURE CAPTIONS
\end{center}
\par\medskip\par

\begin{list}{}{}

\item[{Fig.~1~}]A schematic picture of the precession of orbit and spin
vector, to the leading order in $S$. The vector {\boldmath$J$}
represents the total angular momentum of the particle.
The vector {\boldmath$L$} is orthogonal to the orbital plane and reduces
to the orbital angular momentum in the Newtonian limit.
In the relativistic case, however, these vectors should not be regarded
as well-defined.

\item[{Fig.~2~}]The contours of radii of the inner-most stable circular
orbits on the $(a,S_\perp)$-plane. The conserved angular momentum $J_z$
is assumed to be positive so that the orbits with $a>0$ are co-rotating
with the black hole and those with $a<0$ are counter-rotating.

\end{list}
\end{document}